# GENERATIVE AI IMPACT ON LABOR MARKET: ANALYZING ChatGPT's DEMAND IN JOB ADVERTISEMENTS

Mahdi Ahmadi, Neda Khosh Kheslat, Adebola Akintomide

G Brint Ryan College of Business, University of North Texas

{mahdi.ahmadi, neda.khoshkheslat, adebola.akintomide}@unt.edu

## ABSTRACT

The rapid advancement of Generative AI (Gen AI) technologies, particularly tools like ChatGPT, is significantly impacting the labor market by reshaping job roles and skill requirements. This study examines the demand for ChatGPT-related skills in the U.S. labor market by analyzing job advertisements collected from major job platforms between May and December 2023. Using text mining and topic modeling techniques, we extracted and analyzed the Gen AI-related skills that employers are hiring for. Our analysis identified five distinct ChatGPT-related skill sets: general familiarity, creative content generation, marketing, advanced functionalities (such as prompt engineering), and product development. In addition, the study provides insights into job attributes such as occupation titles, degree requirements, salary ranges, and other relevant job characteristics. These findings highlight the increasing integration of Gen AI across various industries, emphasizing the growing need for both foundational knowledge and advanced technical skills. The study offers valuable insights into the evolving demands of the labor market, as employers seek candidates equipped to leverage generative AI tools to improve productivity, streamline processes, and drive innovation.

## 1. Introduction

The field of Artificial Intelligence (AI) has undergone significant evolution over the past decades, advancing from rule-based systems and early machine learning (ML) models to more sophisticated technologies like Generative AI (Gen AI). Gen AI, particularly through the development of Large Language Models (LLMs), represents a transformative leap in AI capabilities, enabling machines to generate human-like text by leveraging advanced neural network architecture trained on extensive textual data sets (Feuerriegel et al., 2024). Among the most notable advancements in this domain was the introduction of Chat Generative Pre-Trained Transformer (ChatGPT) by OpenAI in November 2022. While ChatGPT is not the first Gen AI tool available to the public, its widespread adoption marks a pivotal moment in AI history, triggering discussions about its potential economic, social, legal, and ethical impacts, particularly on its disruptive influence on the labor market across various industries (Bessen, 2018; Manyika et al., 2017). Understanding how these rapid advancements in Gen AI will reshape the future of work is becoming a critical focus of academic and policy discourse.

Though still in its early stages, Gen AI has already demonstrated a wide range of applications across diverse sectors of business including hospitality, healthcare, finance, and education (Barde & Kulkarni, 2023; Chui et al., 2023; Dwivedi et al., 2024; Hwang & Chen, 2023; P. Zhang & Kamel Boulos, 2023). These applications extend from image recognition and language translation to more complex tasks such as content creation, scenario planning, and automated report generation. As Gen AI becomes further integrated into business processes, it is expected to transform human-machine interactions, enhancing efficiency, productivity, and decision-making at both individual and organizational levels (Brynjolfsson, Mitchell, et al., 2018; Frank et al., 2019; Sturm et al., 2021; West & Allen, 2018). For instance, the integration of Gen AI into customer service has been shown to significantly improve response times and accuracy, while in other fields like computer programming, professional writing, strategic planning, and content creation, it is boosting productivity and enabling more valuable outputs (Al Naqbi et al., 2024; Brynjolfsson et al., 2023; Caner & Bhatti, 2020; Kitsios & Kamariotou, 2021; Noy & Zhang, 2023; Peng et al., 2023).

However, alongside these benefits, there are legitimate legal, ethical, and societal concerns about the broader implications of Gen AI, particularly in relation to job displacement, the widening of skills gaps, and wealth disparity (Acemoglu & Restrepo, 2019; Brynjolfsson & McAfee, 2014; Cazzaniga et al., 2024; Lucchi, 2023) . These dual impacts, promising significant economic benefits while posing serious risks, necessitate a nuanced exploration of how Gen AI will shape the labor market in the coming years. While automation has historically displaced certain job categories, the rise of Gen AI introduces a new level of uncertainty, particularly for knowledge-based professions. There are growing concerns that tasks traditionally seen as requiring human judgment and creativity, such as legal research, journalism, counseling, and software development, could be increasingly performed by AI systems (Frank et al., 2019; Korinek & Stiglitz, 2017). This shift may result in large-scale job displacements, especially in sectors where repetitive or routine cognitive tasks are prevalent. Furthermore, the rapid pace of technological adoption could outstrip the capacity for workforce reskilling, exacerbating existing skills gaps and limiting economic mobility for displaced workers. To better understand these evolving dynamics, our research aims to investigate the effect of Gen AI on the labor market by examining trends in job advertisements. By analyzing the content of AI-related job postings, we seek to uncover emerging patterns that indicate how demand for specific roles and skills may be shifting as Gen AI technologies become more pervasive.

Researchers and experts typically follow one of two approaches to quantify and evaluate the effects of AI on the job market. The first approach is a theory-driven, bottom-up method that begins with a framework to estimate the potential impact of AI on automating tasks currently performed by humans. In this approach, the quantified effects of AI-driven task automation in each occupation are then aggregated to assess broader labor market implications (Brynjolfsson, Mitchell, et al., 2018; Eloundou et al., 2023; Felten et al., 2021; Gmyrek et al., 2023; Webb, 2019).The second approach is a data-driven, top-down method that analyzes job market data to identify the demand for AI-related skills. This method typically involves large-scale collection and analysis of job vacancy data to examine labor market demand at a macro level through textual and contextual analysis (Fareri et al., 2020; Karakatsanis et al., 2017; Pejic-Bach et al., 2020; Pouliakas, 2021; Samek et al., 2021; Squicciarini & Nachtigall, 2021).

Despite the growing body of research on AI's impact on the labor market, there remains a noticeable gap in empirical, data-driven studies that directly examine the evolving demand for AI-related skills. This study addresses that gap by adopting a data-driven approach, combining descriptive statistics and text mining techniques to uncover trends in the U.S. job market related to skills in Gen AI technologies, particularly ChatGPT. We aim to explore key questions such as: How many job positions are being posted in the U.S. that require ChatGPT-related skills? What are the job attributes such as job titles, geographical distribution, salary ranges, and academic degree requirements associated with these roles? Moreover, for what tasks and purposes are employers seeking ChatGPT-related skills? By coupling this analysis with contextual examination of job listings, our results will provide empirical evidence that can be used to evaluate task-level theories and contribute to the expanding body of knowledge on AI's impact on the labor market. The current work aims to offer a unique contribution to the field, providing practical insights into how Gen AI is reshaping job roles and skill requirements across a variety of occupations. The findings of this research will help scholars and experts better understand the demand for Gen AI-related skills and how such demand reflects broader trends in labor market shifts, complementing theoretical frameworks with real-world data.

The remainder of this paper is structured as follows. The next section presents a literature review, where we explore existing models, theories, and frameworks addressing the impact of Gen AI and digital automation on jobs, occupations, and organizations. We also review previous research on the use of natural language processing (NLP) and text mining techniques in analyzing job market demand data. Following this, the research methodology is detailed, outlining our data collection, analysis methods, and tools. In the subsequent section, we present the findings and results of the research, offering insights into how Gen AI is reshaping job roles and skill requirements.

## 2. Literature Review

The term "Generative" in the context of AI refers to the capability of AI systems to recognize and utilize textual, visual, or auditory patterns from extensive datasets to generate new content (Aydin & Karaarslan, 2023; Ellingrud et al., 2023). Gen AI employs deep learning models, such as neural networks and transformers, trained on vast amounts of curated data to produce human-like content in response to complex queries and diverse prompts (Lim et al., 2023). Prominent examples of generative AI include ChatGPT, Claude, GitHub Copilot, and DALL-E, which have garnered widespread attention. Among these, ChatGPT, developed by OpenAI, stands out as a conversational platform powered by a family of advanced natural language models, designed to generate coherent and contextually relevant text-based responses (Felten et al., 2023; Lim et al., 2023).

Numerous scholars and experts contend that the job market is poised for substantial disruption, attributed to the unparalleled advancement in task automation and augmentation by AI technologies. These disruptions are largely driven by productivity gains associated with AI (Brynjolfsson et al., 2023; Brynjolfsson, Rock, et al., 2018; Choudhury et al., 2020; Noy & Zhang, 2023; Peng et al., 2023), as well as transformation in human-machine interactions (Frank et al., 2019; Sturm et al., 2021), and the adoption of new business strategies (Caner & Bhatti, 2020; Kitsios & Kamariotou, 2021). According to an analysis by Goldman Sachs, approximately 75% of jobs in the U.S. and Europe could potentially be automated by AI to some degree, with 25% of those positions at risk of being replaced by Gen AI, and up to 300 million full-time jobs being automated globally (Hatzius, 2023). McKinsey Global Institute further estimates that automation, accelerated by Gen AI, could affect up to 30% of hours worked across various sectors in the U.S. by 2030. Their analysis points to job category shifts, with growth in scientific, technological, and healthcare, while customer service and administrative positions face higher risks of displacement (Ellingrud et al., 2023). A survey conducted by the World Economic Forum (2023) among more than 800 companies across 45 countries found that AI is seen as a key driver of potential structural job displacement, with nearly 75% of surveyed companies planning AI adoption, and 50% anticipating job growth, compared to 25% anticipating job losses due to AI. However, Frey & Osborne (2024) argue that while Gen AI has expanded automation into tasks requiring social intelligence and creativity, it remains limited in unstructured, in-person roles. They conclude that AI is more likely to transform existing jobs than fully replace them, with creative and interpersonal work maintaining a significant role in the labor market.

A common methodology of measuring the impact of AI on jobs is to estimate the effect of AI on the structure of tasks, activities, and occupations. This approach typically begins with a theoretical framework that defines the relationship between AI and tasks, followed by the aggregation of AI exposure metrics at the occupation level to evaluate broader labor market implications. For instance, Felten et al. (2018) introduced the AI Occupational Exposure (AIOE) measure, mapping 10 AI applications to 52 standardized basic human skills. They quantified AI exposure for over 800 standardized occupations, described in the O*NET database of occupations and tasks (Tippins & Hilton, 2010), by aggregating the relatedness of AI applications to human skills. In their revised work, which accounts for advancements in LLMs and Gen AI, they identified telemarketers and various post-secondary teachers, including English teachers, as the most exposed occupations. Furthermore, legal services and securities industries were found to have the highest industrial exposure to LLMs.

Similarly, Brynjolfsson, Mitchell, et al. (2018) developed a "Suitability for Machine Learning" index (SML), which scores suitability of tasks for ML augmentation or replacement. They scored over 18,000 tasks and aggregated them to estimate occupation-level exposure, using the task and occupation data from the O*NET database. Their findings highlighted the significance of AI exposure of occupations like concierges, mechanical drafters, and brokerage clerks. Webb (2019), in a novel approach, created an AI exposure metric by leveraging NLP to analyze the overlap between job tasks from the O*NET database and tasks described in AI-related patents. These task-level scores were aggregated and weighted by their importance in each occupation to generate an overall "exposure" score for how likely AI could automate tasks in that occupation. According to Webb (2019), AI is projected to have the greatest impact on high-skill occupations involving tasks such as pattern recognition, judgment, and optimization. Furthermore, AI is expected to reduce wage inequality between the 90th and 10th percentiles, but it may increase inequality at the top of the wage distribution (Webb, 2019). In another interesting task-based approach, Acemoglu et al. (2022) developed a model by drawing on the three AI exposure indices developed by Felten et al. (2018), Brynjolfsson, Mitchell, et al. (2018) and Webb (2019). They analyzed online job vacancy data from 2010 to 2018 to evaluate AI adoption and its influence on hiring and skill demand. Their findings revealed a significant increase in AI-related vacancies, particularly in establishments with task structures well-suited to AI integration. Also, their findings highlighted that AI adoption led to reduced hiring for non-AI roles, while also shifting demand toward new AI-related skills, rendering some existing skills obsolete (Acemoglu et al., 2022).

Other researchers have conducted variations of task-level AI exposure analysis. Tolan et al. (2021) developed a framework that uses cognitive abilities as an intermediate layer, rather than directly mapping work tasks to AI capabilities. They found that occupations requiring cognitive abilities related to ideas, such as quantitative reasoning, comprehension, and communication, are more likely to be affected by AI advances, whereas jobs relying on physical tasks, like manual labor, are less exposed. Eisfeldt et al. (2023), Eloundou et al. (2023), and Gmyrek et al. (2023) utilized LLMs and text mining techniques to analyze task descriptions and assigned exposure levels to various occupations accordingly. Gmyrek et al. (2023) assessed the global impact of generative AI on employment and found that AI's effect predominantly stems from task augmentation rather than replacement by automation. Additionally, they observed significant disparities in the size of the impact based on gender and between high-income and low-income countries.

Similarly, Cazzaniga et al. (2024) reached comparable conclusions in their multi-country study, modifying Felten et al.'s AIOE index to highlight AI's complementary role in jobs. By factoring in work context and job skills, they distinguished between roles at risk of displacement and those where AI enhances productivity, with advanced economies showing higher exposure but reduced risks for high-skill workers, particularly among higher-income groups. Women and highly educated workers were more exposed across all countries. In another research, Zarifhonarvar (2024) evaluated the impact of ChatGPT on the labor market by combining a supply and demand model and a text-mining analysis of the International Standard Classification of Occupations (ISCO). The study identified which job tasks are most susceptible to AI automation and classified occupations into three categories and finds that up to 70% of jobs could be fully or partially affected by ChatGPT. In another research, Svanberg et al. (2024) introduced a new AI task automation model that assesses both the technical feasibility and economic attractiveness of automating tasks. The study focuses on computer vision tasks and concludes that while AI-driven job displacement is inevitable, its spread will likely be gradual, providing time for policy interventions and retraining efforts.

The second approach to understanding AI's impact on the labor market involves analyzing job market data to identify the demand for AI-related skills and occupations. This data-driven method involves large-scale collection and analysis of job vacancy data, examining labor market demand through textual and contextual analysis. For instance, Squicciarini and Nachtigall (2021) investigated AI-related job postings across Canada, Singapore, the United Kingdom, and the United States from 2012 to 2018, utilizing data from Burning Glass Technologies. AI-related job postings were identified by the presence of at least two AI-related skills from distinct categories, such as AI software, approaches, and applications. The study found a significant increase in AI-related job postings, with a growing number of positions requiring multiple AI skills, particularly in the fields of deep learning and NLP. Additionally, the increasing importance of non-technical skills such as communication, problem-solving, and teamwork suggested that AI-related jobs demand a broader skill set beyond technical expertise (Squicciarini & Nachtigall, 2021). Similarly, Samek et al. (2021) analyzed AI-related job postings in the US and UK using Burning Glass Technologies data from 2012 to 2019. Their text mining and network analysis revealed that Python and Machine Learning were the most in-demand skills, and skill bundles related to AI development, applications, and robotics became more interconnected over time, with neural networks serving as a key link. They also found non-technical skills like communication and teamwork became more prominent, indicating the need for a well-rounded skill set (Samek et al., 2021).

A growing body of research also employs text mining, NLP, ML, and LLMs to analyze job advertisement data and uncover AI-related trends. For example, Bäck et al. (2021) analyzed 480,000 job advertisements from a Finnish online job platform from 2013 to 2020 to explore the demand for AI-related skills. Using a three-tiered glossary that categorized AI-related terms into general AI, core AI technologies, and supporting technologies, they employed text mining and NLP methods to detect AI-related job postings and extract relevant skills. Their study found an increase in demand for specialized AI skills, particularly in the financial and insurance sectors, although the overall number of AI-related postings remained low, indicating early adoption across industries (Bäck et al., 2021). In a similar approach, Kortum et al. (2022) analyzed job advertisements for computer vision and NLP specialists by employing Named Entity Recognition (NER) and term frequency-inverse document frequency (TF-IDF) analysis. They identified distinct skill requirements for computer vision and NLP roles, with computer vision positions focused on hardware-related programming and 3D graphics, whereas NLP roles prioritized Python-based tasks such as sentiment analysis and speech recognition. Both job categories emphasized deep AI and IT skills alongside collaboration and communication, underscoring the diverse skill sets needed for AI-related roles (Kortum et al., 2022). In another study, Verma et al. (2022) performed a content analysis of AI and ML job advertisements in the U.S., scraping job descriptions from Indeed.com and using NLP techniques to map keywords to a predefined skill classification framework. Their findings revealed that machine learning positions primarily required technical skills such as programming and data mining, while AI roles placed greater emphasis on broader occupational skills like communication and decision-making, with a strong geographical concentration of demand in major metropolitan (Verma et al., 2022).

Several studies have used online job posting data without text mining or NLP methods to analyze AI's labor market effects. (Hui et al., 2024; Liu et al., 2023) both employed a difference-in-differences approach to assess the short-term impact of generative AI, particularly ChatGPT, on freelancer employment in AI-exposed fields like writing and programming. Using data from platforms such as Upwork, they found a significant decline in job availability and earnings for freelancers in AI-affected occupations, with a rise in job complexity and competition. Their findings showed top-performing freelancers were disproportionately impacted, and some workers adapted by transitioning from text-related tasks to programming work, suggesting flexibility in response to AI's growing influence. In another research, Alekseeva et al. (2021) analyzed U.S. labor market data from 2010 to 2019 using Burning Glass Technologies and their skill taxonomy to study the demand for AI-related skills. They found a sharp rise in AI skill demand, particularly in large firms with substantial R&D investments, and a notable wage premium for AI-related roles. Interestingly, firms that demanded AI skills also tended to offer higher wages for non-AI positions, indicating a broader impact on the labor market (Alekseeva et al., 2021).

Beyond the impact of AI on employment, text mining, NLP, and ML have been applied to job advertisement data in other areas. For instance, Verma et al. (2023) analyzed Indeed.com postings to develop a skill framework for augmented and virtual reality positions, identifying high demand for user interface and experience design, programming, and graphics rendering. In a different study, Pejic-Bach et al. (2020) used LinkedIn data to highlight key competencies for Industry 4.0 roles, including cyber-physical systems and Internet of Things, as well as general business skills like supply chain management. With a focus on education, Benhayoun and Lang (2021) examined gaps between AI-related training in French schools and AI job market demands, revealing deficiencies in certifications and interdisciplinary skills. In a similar paper, Karakolis et al. (2022) developed a system using text-mining to align university curricula with labor market needs, highlighting a persistent gap between education and industry requirements.

Although research on the impact of AI on the job market is expanding, the field is still developing, and more research is needed to deepen our understanding. Frank et al. (2019) identified several barriers that hinder efforts to measure AI's effects on the future of work. These include the lack of high-quality data on the dynamic nature of occupations, as well as a lack of empirically informed models for understanding key processes such as skill augmentation and replacement. Overcoming these barriers requires better and more data collection and analysis. In our study, we address some of these challenges by analyzing job demand data and extracting AI-related skills and requirements from job advertisements. By examining labor market trends, our research contributes to the evolving understanding of AI's impact on the workforce.

## 3. Methodology

In this research we address the following questions: How many job opening advertisements with ChatGPT as a demanded skill were posted online in the U.S. during the research period? What are the posted job attributes such as job titles, geographical distribution, salary ranges, contract type, and degree requirement? What are the required skills, academic degrees, and years of experience? And for what tasks and purposes were employers demanding of the job applicants to possess ChatGPT-related skills? To address these questions, we collected job advertisement data from popular job posting web platforms and conducted text mining and NLP analysis. Our research methodology is outlined in Figure 1.

In this study, we followed a procedure to identify, screen, and include job postings as they seem relevant to the research questions. We collected data from three popular job posting websites: Indeed, LinkedIn, and ZipRecruiter. We used the term "ChatGPT" and its variation to retrieve relevant postings. Data collection was done weekly between May and December 2023. All textual elements of each posting were collected and stored in separate documents, which were used to build the corpus for textual analysis. In the collected documents there were a few instances of irrelevant job postings that were retrieved because they contained the term "ChatGPT". For instance, some employers stated the use of ChatGPT is forbidden by job applicants for preparing their resume or cover letters, and few other employers mentioned they used

ChatGPT to write the job posting. After manually checking a sample of data, we used regular expressions to identify and remove such irrelevant documents. In addition, there were many duplicate postings across the three sources. We found and removed duplicates by applying TF-IDF to convert each document into a sparse matrix using cosine similarity as the distance function (Salton & McGill, 1983). After data cleaning and integration, our corpus includes 1128 unique job posting documents. The data set is provided in the Supplementary Materials.

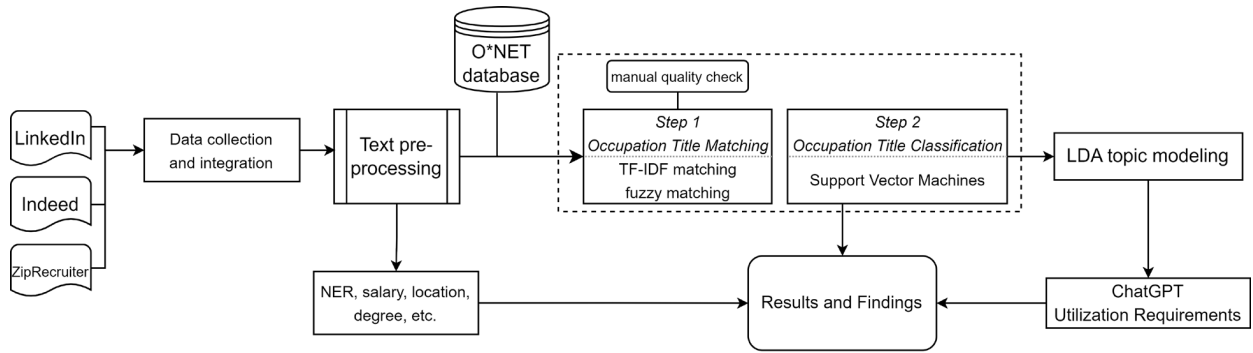

**Figure 1.** Research methodology steps

## 3.1. Data Collection and Pre-Processing

The corpus of job posting documents were semi-structured. Our web scraping process allowed us to tag certain textual elements during the data collection such as "job title", "location", "job description", "employer name", "salary range", and "contract type". To further structure the data and extract more information, we performed keyword matching, regular expression search, and Named Entity Recognition (NER) on the documents. NER involves identifying and classifying named entities in a text into predefined categories such as skills, platforms, organizations, locations, degrees, and other significant entities (Nadeau & Sekine, 2007). This process enhances the semantic understanding of text data, facilitating accurate information retrieval in NLP applications, and it has been widely employed in job market data analysis (Gaur et al., 2021; Upadhyay et al., 2021; Zhao et al., 2015).

## 3.2. Job Title Identification

An important task in studying job posting data systematically is to identify and classify postings based on the job titles, which are written in natural language. Given the complexity of natural language data, NLP and ML techniques are commonly applied (Javed et al., 2016; Liu et al., 2022; Nasser & Alzaanin, 2020;

Rahhal et al., 2023). We used a two-step process to identify job titles, as shown in Figure 1. In both steps we use the Occupational Information Network (O*NET) database as our occupation structure reference. O*NET was developed in the mid-1990s with sponsorship from the U.S. Department of Labor, provides comprehensive and standardized descriptors for approximately 1,000 occupations across the U.S. economy. The database includes standardized occupation titles and itemized data on job requirements and worker attributes, and it provides a three-level structure for categorizing occupations. Its utility in labor market data analysis has been well-documented in various scholarly works and practical applications (Burrus et al., 2013; Peterson et al., 2001; Tippins & Hilton, 2010), and it has been cross-referenced with other international occupation databases such as International Standard Classification of Occupations (ISCO) (Hardy et al., 2018) and European Skills, Competences, Qualifications and Occupations (ESCO) (The Crosswalk between ESCO and O*NET, 2022).

To map job vacancy titles to the closest O*NET occupation titles, we first applied a cosine similarity search on the TF-IDF matrices of the job titles and ONET occupation titles. We set the threshold of the similarity score to 0.95 to match strings with very high degree of similarity. Next, we applied a partial ratio fuzzy string search using the Fuzzy-Wuzzy Python library (Cohen, 2011), based on the Levenshtein distance (Navarro, 2001) between the job titles in the advertisements and O*NET occupation titles. We used a 0.95 threshold for the Levenshtein distance to ensure highly similar string matching. Through these two methods, we found ONET occupation matches for nearly 90% of the job postings. All matches were manually checked to ensure quality and accuracy.

After matching 90% of job posting to O*NET occupation titles, we used these matches to train a classification model. For this, the job description text data was cleaned through stop words removal and lemmatization, and then vectorized using TF-IDF to convert text into numerical features. We trained a linear kernel Support Vector Machine (SVM) classification model, using the O*NET occupation title as the target variable, and employing a one-versus-rest strategy to handle the multi-class classification problem. SVM is a supervised learning algorithm that aims to find the optimal hyperplane that best separates different classes of data (Cortes & Vapnik, 1995; Cristianini & Shawe-Taylor, 2000). SVM is particularly effective for high-dimensional data, such as text (Joachims, 1998; Tong & Koller, 2001; W. Zhang et al., 2008), and it is widely used in labor market data classification (Goindani et al., 2017; Javed et al., 2015, 2016; Rahhal et al., 2023). Our tunned SVM model achieved a false positive rate of 0.5%, with precision and accuracy rates of 99%. We ensured model robustness by performing 5-fold cross-validation during training. The model was then applied to the remaining job postings to assign O*NET occupation titles. We manually cross-checked the model output for quality and accuracy. Once all records were assigned a standard occupation title, we used the O*NET structure to classify each job posting into Occupation Family Levels 1, 2, and 3.

### 3.3. Topic Modeling

To analyze the specific ChatGPT skills employers require, we employed Latent Dirichlet Allocation (LDA) to extract relevant information from job advertisement data. LDA is a statistical method that models word distributions within a corpus to uncover latent structures, known as topics, making it a powerful tool for text analysis (Blei et al., 2003). In this study, we isolated the segments of job descriptions that specifically referred to ChatGPT by extracting the sentences containing and surrounding the term. We then applied LDA to this curated data set to identify recurring themes and topics related to ChatGPT skills and functions. The output of the LDA model included clusters of words and terms that captured the variability within the data, effectively summarizing the key information from each document. Each word cluster was labeled to describe the ChatGPT-related skills or utility requirements.

We optimized the performance of the LDA model by tuning its hyperparameters, evaluating the model quality using coherence and perplexity metrics. Coherence measures the degree of semantic similarity between words within a topic, reflecting how interpretable and meaningful the topics are (Mimno et al., 2011), while perplexity assesses the model's ability to generalize to unseen data, with lower values indicating better performance (Blei et al., 2003). In addition to these metrics, we manually examined the output word clusters to ensure the topics were both statistically sound and easily interpretable.

Finally, we mapped the identified ChatGPT-related skills and requirements topics to the O*NET occupation families established in the previous step. Following the method proposed by De Mauro et al. (2018), we calculated the relevance of each ChatGPT-related skill set to the occupation families. We used the document-topic distribution matrix (Θ) of the LDA model output, where each row represents a document (i.e., a job ad) and each column represents a topic (i.e., a ChatGPT skill set). The values in this matrix represent the probability of each topic within each document. We averaged the topic probabilities by the identified O*NET occupation family level 1. The resulting matrix A contains the average probability of existence of a ChatGPT-related skill set within each occupation family. We normalized A by dividing each element by its row-wise average to obtain a matrix $\hat{A}$ which represents the importance of each ChatGPT skill set $s$ in an O*NET occupation family $f$ compared to other ChatGPT-related skill sets. To visualize the results, we used the following discretization schema:

$$\hat{A}_{f,s} \leq 0.25 : \text{ChatGPT skill set } s \text{ is } not\ central \text{ to jobs in family } f, \square\square\square$$

$$0.25 < \hat{A}_{f,s} \leq 0.5 : \text{ChatGPT skill set } s \text{ is } slightly\ central \text{ to jobs in family } f, \blacksquare\square\square$$

$0.5 < \hat{A}_{f,s} \leq 7.5$ : ChatGPT skill set $s$ is *moderatly central* to jobs in family $f$, ■■□

$0.75 < \hat{A}_{f,s}$ : ChatGPT skill set $s$ is *highly central* to jobs in family $f$, ■■■

## 4. Results and Discussion

### 4.1. Job Trends and Attributes

Figure 2 illustrates the trends in ChatGPT-related job postings across nine different O*NET level 1 occupation families, where 69 unique occupation titles were identified. The "Computer and Mathematical" family leads in demand, with "Software Developers" representing nearly 20% of all job ads. Notably, marketing-related roles occupy three of the top five demanded occupations, including "Marketing Managers," "Market Research Analysts and Marketing Specialists," and "Search Marketing Strategists." These roles emphasize the importance of generative AI in content generation and management, search engine optimization (SEO), and brand strategy, underscoring the usability of large language models (LLMs) in marketing activities. Additionally, "Writers" has also emerged as a significant occupation, reflecting the growing role of LLMs in generating textual content, which supports writers, editors, and content specialists in automating and enhancing their creative processes.

In contrast, the "Legal" occupation family has the least representation, with only eight job postings, primarily seeking paralegals and attorneys knowledgeable in ChatGPT and Gen AI. Although still a niche labor market, there are clear signs of growth. Our follow-up analysis of job postings shows an increasing number of ads requesting legal experts with generative AI skills. The applications of generative AI, particularly LLMs, are beginning to significantly impact legal and law-related occupations by automating routine tasks, enhancing legal research, improving decision-making processes, drafting contracts, summarizing legal documents, and conducting complex case law analysis. This automation allows legal professionals to focus on the more complex aspects of their work, aligning with the findings from other researchers (Hu & Lu, 2019; Legg & Bell, 2019; Surden, 2019; Waisberg & Hudek, 2021). Despite concerns over the ethical use of these technologies, particularly regarding biases in AI-generated legal advice, the field is expanding rapidly. The business potential in this area is promising, as several companies are developing LLMs fine-tuned on legal content to offer customized services. This is reflected in the number of job ads we observed in our follow-up searches, which increasingly demand specialists with dual expertise in law and generative AI development. New roles, such as legal technologists, are also emerging, collaborating with both technology developers and attorneys to streamline workflows, automate document processes, and provide customized AI integrations that help legal teams save time, reduce costs, and better serve clients.

Compared to IT, business, and management jobs, education-related positions are still a small share with 2.3% of the job advertisements. However, we found a noticeable number and variety of job vacancies under the "Educational Instruction and Library" family, including roles such as business instructors, math and programming teachers, and career training coaches. These positions often involve two distinct responsibilities: teaching principles and functions of LLMs, ML, NLP, and Gen AI, or utilizing tools like ChatGPT to create educational materials for other subjects. There is a growing body of research showing the transformative potential of Gen AI in education. AI can personalize learning experiences, automate administrative tasks, offer real-time feedback, and assist educators in grading and curriculum development, as well as provide intelligent tutoring systems that enhance educational efficiency and effectiveness (Dempere et al., 2023; Hwang & Chen, 2023; Lim et al., 2023; Mollick & Mollick, 2023; Rahman & Watanobe, 2023). Our follow-up job vacancy search reveals a rising trend in hiring educators with varying degrees of familiarity with ChatGPT and other Gen AI tools, signaling the increasing demand for integration of AI into educational practice.

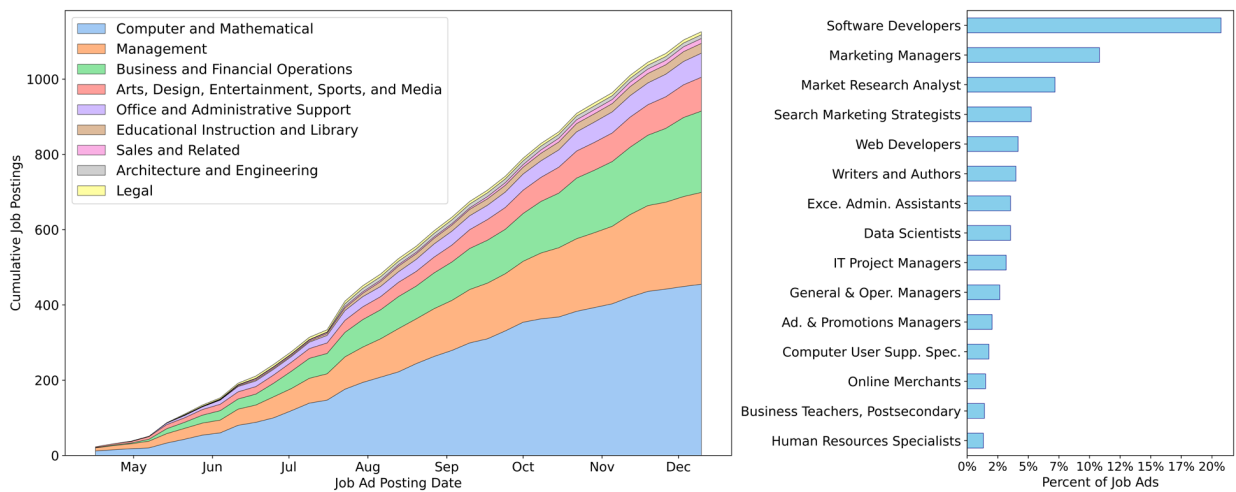

**Figure 2.** ChatGPT related job posting trend by O*NET Family Level 1 (left) and top O*NET occupation titles

The findings from the analysis of job postings, as illustrated in Figure 3, indicate that nearly 50% of the job ads originate from employers in four states: California, Texas, New York, and Florida (Figure 3a). The majority of these postings are for full-time positions, with over 38% offering remote work options, reflecting the noticeable flexibility in employment conditions for roles requiring Gen AI skills. Annual salaries for these jobs vary significantly depending on job title, location, and required experience. Notably, occupations in the "Computer & Mathematical," "Architecture and Engineering," and "Sales and Related" families offer the highest average salaries, while roles in "Office and Administrative Support," such as Executive Secretaries and Executive Administrative Assistants, tend to offer the lowest salaries on average (Figure 3b). For these administrative positions, the preliminary requirement related to ChatGPT is familiarity with the tool and its use in daily operational and administrative tasks.

In terms of experience requirements, the majority of job postings require less than five years of experience, with fewer than 10% requiring more than ten years, primarily for senior software developer or AI development project manager roles (Figure 3c). Regarding educational qualifications, more than 65% of job ads specify a bachelor's degree as the minimum requirement, with only 9.4% preferring or requiring a PhD, predominantly for research-intensive and highly technical positions in data science and AI development (Figure 3d). These trends suggest that while advanced skills and degrees are in demand, most job openings focus on mid-level experience and practical knowledge of AI and related technologies.

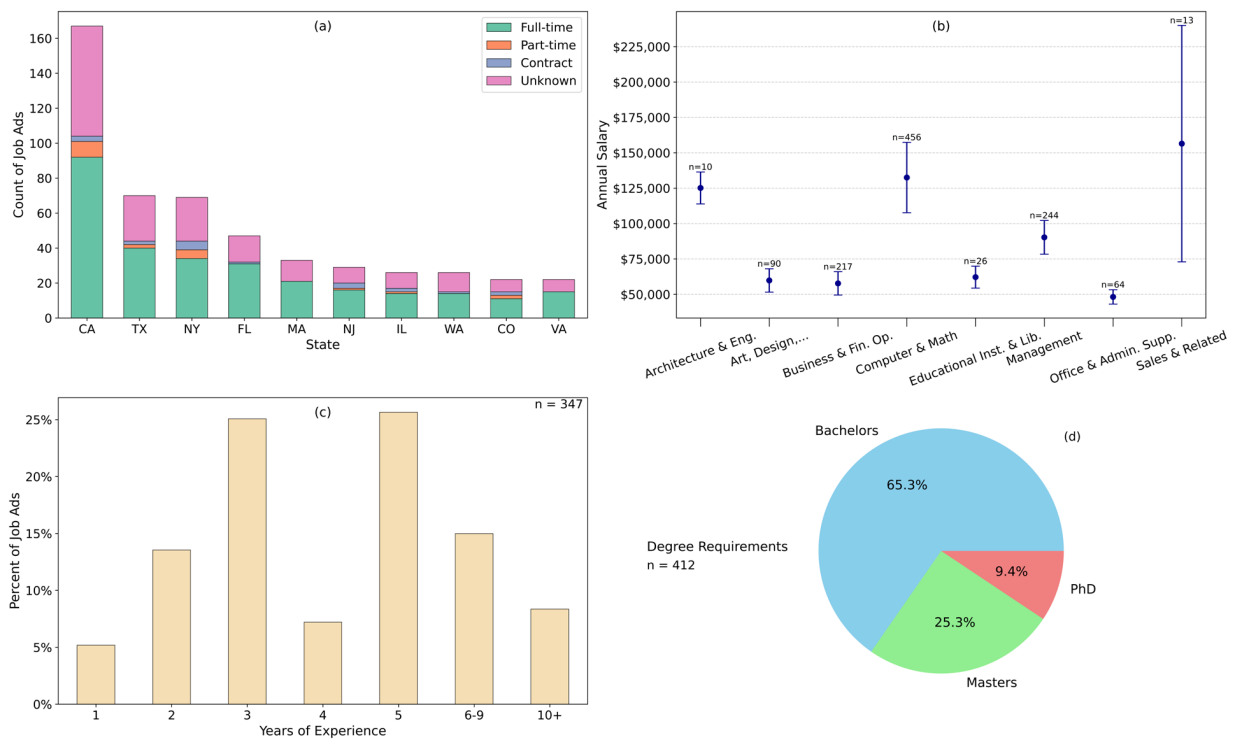

**Figure 3.** Attributes of ChatGPT related posted jobs

## 4.2. Gen AI Skills

In the topic modeling of job advertisements content related to ChatGPT skills, five distinct topics were identified. The first and most prevalent topic, labeled "General Familiarity" (skill set #1), accounts for almost 42% of the job ads. This category reflects cases where employers expect candidates to be generally familiar with ChatGPT or other Gen AI tools. These job ads do not specify particular tasks but emphasize the importance of candidates knowing how to use Gen AI technologies to improve productivity, enhance the speed and accuracy of work, summarize documents, answer questions, and bring efficiency to repetitive tasks. Familiarity with Gen AI tools is often mentioned as a plus, with candidates expected to utilize them both individually and in team settings to enhance their performance. As shown in Table 1, the label for this skill set was derived from the most probable keywords (e.g., "familiarity," "productivity," "tool") that emerged from the LDA analysis, which we further refined through manual examination of the job ads.

**Table 1.** The 15 most probable words associated with each ChatGPT skill set, per LDA output. The labels reflect human interpretation of the overall theme of each skill set.

| Skillset # | 1 | 2 | 3 | 4 | 5 |
|---|---|---|---|---|---|
| Label | *General Familiarity* | *Creative Content Generation* | *Marketing* | *Advanced Functionalities* | *AI Product Development* |
| % of Job Ads | 41.8% | 14.5% | 12.7% | 16.3% | 14.8% |
| Top keywords | familiarity | content | marketing | coding | llms |
| | work | writing | seo | prompt | solution |
| | tool | creation | branding | evaluate | product |
| | experience | email | crm | python | ml |
| | technology | proficiency | email | data | nlp |
| | productivity | adobe | customer | knowledge | model |
| | generative | social media | strategy | optimize | fine-tuning |
| | skill | research | business | azure | design |
| | ability | strong | campaign | automating | integration |
| | team | video | communication | BERT | application |
| | plus | editing | sales | analysis | technique |
| | new | drafting | conversion | platform | commercialize |
| | management | project | networking | microsoft | build |
| | similar | engaging | content | sql | algorithms |
| | google | summarizing | social media | performance | adaptation |

The second skill set, labeled as "Creative Content Generation", is the primary focus in 14.5% of the job ads. Under this skill set, employers explicitly request candidates to be able to generate textual and visual content using Gen AI tools such as ChatGPT, DALL-E, MidJourney, Stable Diffusion, Adobe Firefly, Claude, and Google Gemini. The demands within this topic are diverse, ranging from writing and editing content for emails, blog posts, and presentations to producing creative scripts and social media captions. Employers also expect candidates to have capability of employing generative AI tools for tasks such as accelerating content development, enhancing the quality and effectiveness of copy, and managing social media platforms. Additionally, the ability to develop engaging content for websites and improve user experience is often highlighted, emphasizing the growing role of AI in content ideation and production for users.

The third identified topic, labeled "Marketing" (skill set #3), appears in almost 13% of the job ads. The demand for this skill set targets candidates who can leverage Gen AI tools for generating marketing, promotional, and customer engagement content. Employers are looking for candidates capable of using AI to assist with a wide range of marketing tasks, including email marketing, branding, networking with potential customers, and enhancing sales strategies. These roles also involve planning and creating SEO strategies, optimizing keywords, and developing marketing campaigns. Notably, a significant number of job ads specifically mention the need for experience using ChatGPT for SEO and keyword optimization, highlighting the growing importance of AI in improving web traffic conversions. In addition to AI skills, candidates are expected to combine their ChatGPT experience with expertise in customer relationship management (CRM) systems and sales platforms to develop and analyze marketing strategies. As reflected in the keywords in Table 1, such as "marketing," "seo," "branding," and "crm", these roles require a blend of traditional marketing skills and generative AI capabilities, emphasizing the need for proficiency in both areas.

**Table 2.** Degree of centrality of ChatGPT skill sets across the identified O*NET Occupation Families.

| O*NET Occupation Family Level 1 | Job Ads | General Familiarity | Creative Content Generation | Marketing | Advanced Functionalities | AI Product Development |
|---|---|---|---|---|---|---|
| Architecture and Engineering | 10 | ■■■ | □□□ | □□□ | □□□ | □□□ |
| Arts, Design, Entertainment, Sports, and Media | 90 | ■■□ | ■■■ | ■□□ | □□□ | □□□ |
| Business and Financial Operations | 217 | ■■■ | ■□□ | ■■□ | ■□□ | □□□ |
| Computer and Mathematical | 456 | ■□□ | □□□ | □□□ | ■■■ | ■■■ |
| Educational Instruction and Library | 26 | ■■■ | ■■□ | □□□ | □□□ | □□□ |
| Legal | 8 | ■■■ | ■□□ | □□□ | ■□□ | □□□ |
| Management | 244 | ■■■ | ■□□ | ■■□ | ■□□ | □□□ |
| Office and Administrative Support | 64 | ■■■ | ■□□ | ■■□ | □□□ | □□□ |
| Sales and Related | 13 | ■■□ | □□□ | ■■■ | □□□ | □□□ |

The fourth skill set, labeled "Advanced Functionality", is found in 16.3% of the job ads. This category of the skill demand focuses on job applicants who are expected to have a deeper understanding of generative AI tools like ChatGPT and be skilled in prompt engineering. Prompt engineering refers to the practice of designing and refining the prompts given to Gen AI models in order to generate the most accurate, relevant, and useful responses. Job applicants in this category are required to not only know how to create effective prompts but also optimize strategies for generating them and evaluate the outputs of Gen AI models, continuously improving them using prompt templates, feedback loops, and systematic design techniques. Another key skill in this category is the ability to use Gen AI tools, such as ChatGPT and GitHub Copilot, to assist in coding tasks. Candidates are expected to write and optimize code with the help of these LLMs, enabling faster and more efficient software development processes. Additionally, employers look for applicants with extensive experience in working with conventional development tools such as SQL, Azure, AWS, and other cloud-based technologies. The ability to integrate Gen AI capabilities into these platforms to enhance productivity, reduce costs, and increase overall efficiency is critical for roles that demand these advanced functionalities. As shown in Table 1, keywords like "coding," "prompt," "evaluate," "optimize," and "azure" highlight the focus on combining advanced Gen AI capabilities with broader development tools to meet organizational goals.

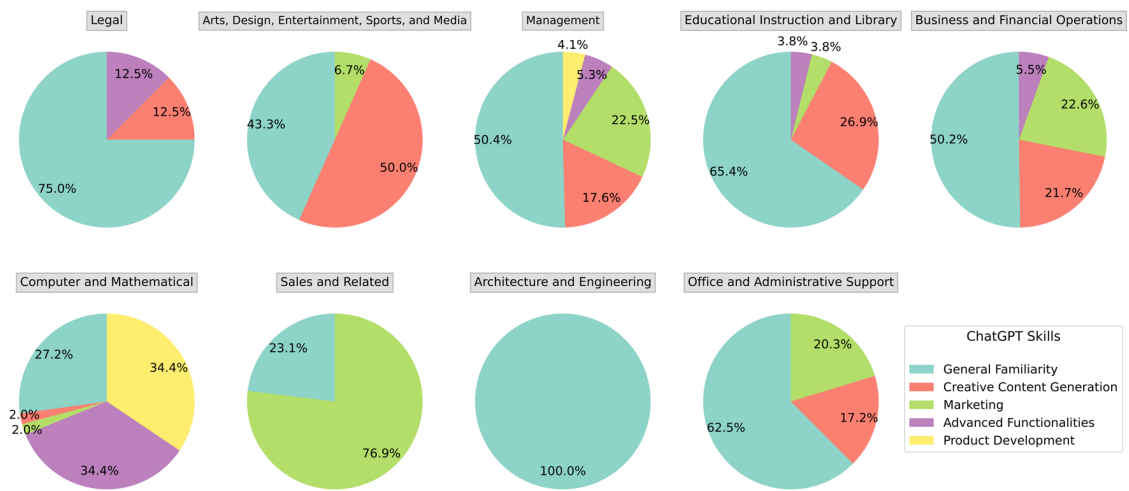

**Figure 4.** Distribution of ChatGPT skill requirements across the identified O*NET Occupation Families.

The last and most advanced skill set identified in our analysis is labeled "Product Development" (skill set #5 in Table 1), comprising almost 15% of the job advertisements. Employers in this category are looking for applicants who possess the skills to develop and commercialize Gen AI products, specifically focusing on customizing and producing LLMs like the GPT family models for business applications. Candidates are expected to have the expertise to integrate LLMs into various products such as chatbots, document management systems, review platforms, project planning tools, and other internal or customer-facing solutions. This includes integrating Gen AI with conventional platforms like Azure, Amazon Web Services, Google Cloud Platform, as well as other backend systems, or creating Gen AI browser plugins that streamline user experience and product functionality. Additionally, applicants are required to demonstrate proficiency in architecting, developing, training, and fine-tuning LLMs similar to ChatGPT, ensuring these models are optimized for specific business needs. Another important demand in this category is the ability to conduct research and implement cutting-edge NLP and ML techniques to enhance the performance and accuracy of AI-driven products. This is particularly crucial in solving complex business problems through AI-powered solutions.

Moreover, high-level skills in coding, algorithm design, software development, system integration, and working with APIs are essential for roles in this category. As reflected in the keywords from Table 1, terms like "llms," "product," "fine-tuning," "ml," and "integration" emphasize the advanced technical expertise required to bring Gen AI products to market, solve business challenges, and drive innovation. Employers seeking candidates with the "Product Development" skill set are not just users of generative AI tools like ChatGPT; their core business revolves around developing and enhancing such technologies.

The analysis of skill set centrality across various occupation families provides insightful patterns about the demand for ChatGPT-related skills. As shown in Table 2, General Familiarity with ChatGPT and Gen AI tools is central to all occupation families to varying degrees. This suggests that employers across industries are increasingly seeking candidates with basic knowledge of these tools, reflecting their importance for enhancing productivity, streamlining tasks, and optimizing workflows. Figure 4 supports this, showing that the share of job ads emphasizing General Familiarity ranges from 23% in Sales to as high as 100% in Architecture and Engineering, indicating that basic Gen AI proficiency has become a common requirement across diverse sectors.

In the Computer and Mathematical occupation family, skill sets such as Advanced Functionalities and Product Development are moderately central, as indicated in Table 2. This aligns with the technical nature of tasks and responsibilities in this family, which often involve the integration, advanced applications, and development of AI tools. The share of these two skill sets is substantial in Figure 4, with Advanced Functionalities and Product Development each accounting for 34.4% of the job ads. These roles typically require expertise in coding, model fine-tuning, and the development of generative AI products, which are integral to the advancement of AI technologies. Notably, the Management occupation family is the only other category where Product Development is a key demanded Gen AI skill set. Although significantly fewer job ads were observed, they were specifically for manager positions in software development, requiring candidates to have hands-on experience and deep knowledge of how LLMs are developed and optimized. This indicates that for these managerial roles, a strong technical foundation is crucial for overseeing AI product development.

In contrast, for occupation families like Office and Administrative Support and Educational Instruction and Library, the demand for more advanced Gen AI skills is lower. In the Office and Administrative Support family, for example, the General Familiarity skill set is the primary focus in over 60% of the job ads, while Creative Content Generation and Marketing play moderate roles. This reflects the typical tasks in this family, such as handling business communications and presentations, automating office tasks, answering business questions, and summarizing documents. However, Advanced Functionalities and Product Development are either slightly central or not central at all, indicating that these positions do not require deep technical expertise, instead focusing on basic uses of Gen AI tools to improve everyday workflows and productivity.

The Management and Business and Financial Operations families exhibit a similar distribution of Gen AI skill sets, as shown in Table 2 and Figure 4. While General Familiarity with ChatGPT and similar technologies is highly central to both, the second most sought-after skill set is experience and knowledge in performing marketing tasks using such tools. A close contextual examination of job ads in these occupation families shows that employers expect candidates to possess Gen AI skills alongside conventional experience in marketing, sales, and customer management tools and frameworks. This suggests that the need for augmented and integrated skill sets is growing as Gen AI tools become increasingly capable of performing specialized tasks. In both families, the third most in-demand skill set is Creative Content Generation, with a focus on report generation, document summarization, business research compilation, website content production, search engine optimization, and communication automation. A small number of jobs in these occupation families also demand advanced Gen AI skills, specifically for tasks such as using ChatGPT and other Gen AI tools to write scripts for automating repetitive tasks, generating SQL queries, and utilizing these tools within cloud environments like Azure, AWS, and GCP.

Overall, the varying degrees of centrality and the distribution of skill sets across different occupation families highlight the diverse nature of responsibilities and tasks in these sectors. Occupations requiring higher technical expertise, such as those in Computer and Mathematical roles, demand more advanced AI skills, while administrative and support roles primarily focus on basic familiarity with AI tools.

## 5. Conclusion

This study examined the growing demand for ChatGPT-related skills in the U.S. labor market, revealing the extent to which Gen AI technologies such as ChatGPT are becoming essential across various sectors. By analyzing job advertisements from multiple platforms, we identified five primary skill sets employers seek: general familiarity, creative content generation, marketing, advanced functionalities, and product development. The findings indicate that "General Familiarity" with ChatGPT is the most demanded skill set across multiple industries, reflecting the importance of basic AI proficiency in today's job market.

The results demonstrate how Gen AI tools like ChatGPT are reshaping job roles beyond traditional tech-centric fields. Industries such as marketing, administrative support, and education are incorporating AI tools to streamline tasks and enhance productivity. Notably, advanced functionalities and product development skills were central in highly technical roles, particularly within the Computer and Mathematical occupation family. The emergence of roles focused on developing AI products signals a growing market for specialized AI developers and engineers, highlighting the increasing importance of technical expertise in AI development.

Beyond technical tasks, this research underscores the broader integration of AI skills across various sectors. Job roles are evolving as Gen AI tools like ChatGPT are used for decision support, content creation, and business management. These findings point to the need for diverse skill sets, blending traditional expertise with AI capabilities. This shift has important implications for workforce development and highlights the need for professional training programs. Furthermore, higher education institutions must integrate these skills into their curricula and revise their teaching materials to prepare students for the rapidly changing job market. Incorporating AI-related skills into various disciplines, from business to engineering, will be crucial in ensuring that graduates are equipped for future roles.

Finally, this research contributes to a deeper understanding of the evolving job market, offering valuable insights into how generative AI is influencing labor demands. The study's findings provide a data-driven foundation for predicting future job trends and can inform labor market models that account for the increasing importance of AI-related skills. Furthermore, our findings are helpful for understanding how tasks are augmented or replaced with Gen AI technologies. Future research should focus on more data-driven studies that continuously monitor and analyze job market demand as AI technologies evolve. Additionally, empirical studies are needed to explore how much and in what ways Gen AI tools are being used by workers in various roles. It would also be valuable to examine the long-term impact of AI on job retention and displacement, as well as the specific skills and training required to bridge any gaps between AI proficiency and practical job performance.